\documentclass[aps,twocolumn,showpacs]{revtex4}

\usepackage{graphicx}

\def\be{\begin{equation}}
\def\ee{\end{equation}}
\def\bc{\begin{center}}
\def\ec{\end{center}}
\def\lan{\langle}
\def\ran{\rangle}

\begin{document}

\title{Pacman percolation: a model for enzyme gel degradation}
\author{T. Abete$^1$, A. de Candia$^1$, D. Lairez$^2$, A. Coniglio$^1$}
\affiliation{$^1$Dipartimento di Scienze Fisiche, Universit\`{a} di Napoli
``Federico II'' and INFM, Unit\`{a} di Napoli\\
Complesso Universitario di Monte S. Angelo, via Cintia, 80126 Napoli, Italy\\
$^2$ Laboratoire L\'{e}on Brillouin, CEA-CNRS, CEA/Saclay, Saclay, France}


\begin{abstract}
We study a model for the gel degradation by an enzyme,
where the gel is schematized
as a cubic lattice,
and the enzyme as a random
walker, that cuts the bonds over which it passes. The model undergoes a
(reverse) percolation transition, which for low density of enzymes falls
in a universality class different from random percolation. In particular
we have measured a gel fraction critical exponent $\beta=1.0\pm 0.1$,
in excellent agreement with experiments made on the real system.
\end{abstract}

\maketitle

The extracellular matrix (ECM) is a gel composed by various proteins,
including collagen, elastin, fibronectin and laminin,
connected to form an elastic network that extends macroscopically.
This gel is normally
impermeable to cell passage, and ensures organ integrity by insulating
organs and preventing cell dissemination. Moreover it is the support of
cell adhesion and regulates cell proliferation, differentiation and
locomotion.
During specific processes, the ECM can be degraded
by a variety of proteolytic enzymes, especially metalloproteinases,
that catalyze the hydrolysis of the peptide bonds,
increasing the permeability of the ECM to the passage of
cells. This degradation process can at some point solubilize the gel,
realizing a reverse ``gel-sol'' transition, and bringing the ECM
to a liquid state, in which cells are no longer confined and can
freely diffuse.  This solubilizing transition is especially connected
with tumor invasiveness, in which some cells
access the lymphatic and blood circulation, and disseminate to distant
organs (metastasis). In this view, beyond the biochemical processes
involved at molecular level, the understanding of the physical mechanisms of
the ECM degradation is of great importance.

The passage from a liquid to a gel state, is a critical
phenomenon \cite{stauffer,degennes}, in which soluble monomers bind to
form larger and larger clusters. At some point, when the bond density $p$
becomes grater than some critical threshold $p_c$, an ``infinite'' cluster
extending macroscopically is formed, and the network becomes a solid gel
with an elastic response. The reverse transition, in which bonds are
removed, and the system goes from a gel to a liquid state,
can be clearly described in the same framework.
The theory of critical phenomena predicts
that, near the transition, the macroscopic quantities describing the system
are related to the distance from the transition $(p-p_c)$ by power laws. The
average cluster diameter diverges as $|p-p_c|^{-\nu}$, while the weight 
average mass as $|p-p_c|^{-\gamma}$. The viscosity diverges as $|p-p_c|^{-k}$
below the transition ($p<p_c$), and stays infinite above it, while
the gel fraction (the density of the infinite cluster)
and the elastic modulus, that are zero below the transition,
grow above it respectively as $|p-p_c|^{\beta}$ and $|p-p_c|^{t}$.
It is important to point out that exponents $\nu$, $\beta$, $\gamma$,
etc\ldots
are universal, that is they do not depend on the microscopic details of
the system, but only on characteristics like the dimensionality
of the system, or whether or not there is a long range correlation in the
distribution of the bonds.

Recently, a series of interesting experiments have been realized to study the
{\em in vitro} degradation of protein gels by exogenous proteinases,
under cell-free conditions \cite{BBA2000,BPJ2003}.
In particular, in \cite{BBA2000} it was shown that a gel-sol
transition adequately describes the degradation of the gel.
Two kinds of gel, fibronectin and ECM gel, and three kinds of
enzyme, thermolysin, trypsin and proteinase K, were used in various
combinations. An enzyme solution was added to a certain quantity of gel,
and the solubilized fraction $X_{\text{sol}}(t)$ of peptides was
measured as a function of time. The balance between enzyme diffusion
inside the matrix, and proteolysis reaction, was
such that the enzyme reached an homogeneous distribution inside the gel
before the reaction had proceeded to a notable extent, so that the
experiments studied the volume, and not surface, degradation of the gel.
The gel-sol transition is reached at some time $t_c$,
when $X_{\text{sol}}(t)$ becomes equal to one, or when
the gel fraction $X_{\text{gel}}(t)=1-X_{\text{sol}}(t)$ becomes zero.
With various gel-enzyme combinations, and different enzyme
concentrations, it was found that
$X_{\text{gel}}\propto|t-t_c|^\beta$, with $\beta\simeq 1$, for
$t<t_c$. For ECM gel and trypsin for example $\beta=1.01\pm 0.03$
\cite{BBA2000}. It may be assumed that the density of hydrolyzed bonds
is a linear function of time,$(p-p_c)\propto (t-t_c)$, at least near
the transition threshold, so that
$X_{\text{gel}}\propto|p-p_c|^\beta$, with $\beta\simeq 1$, for $p>p_c$.

This result is quite unexpected, because sol-gel transition is usually
well described by random percolation, which is obtained when each
bond between two monomers is present with probability $p$,
and there is no correlation between different bonds. Random percolation
in three dimensions gives a critical exponent $\beta=0.41$, very different
from the one measured in the gel degradation experiments.
In Ref.\ \cite{BBA2000} different possible explanations of this discrepancy
are proposed, including the fact that the proteins of the gel may be
considered as interpenetrated polymer coils, that not all the solubilized
fraction may be measured in the experiments, or that bonds of the
gel phase may be catalyzed more efficiently than those of the liquid phase.

Another possibility, in order to explain the change in the
universality class with respect to random percolation, is the presence
of a long range correlation in the distribution of non-hydrolyzed
bonds \cite{BPJ2003}. The correlation function is defined as $G(|{\bf
r}|)=\lan\rho({\bf r}^\prime)\rho({\bf r}^\prime+{\bf r})\ran
-\lan\rho({\bf r}^\prime)\ran^2$, where $\rho({\bf r})$
is the density of bonds, and the average $\lan{\cdot}\ran$ is done over
the reference position ${\bf r}^\prime$.
It was shown by Weinrib and Halperin \cite{weinrib} that if the correlation
obeys a power law $G(r)\simeq r^{-a}$ at long distances with $a<d$ (where
$d$ is the dimensionality of the system), then the percolation transition
falls in a universality class different from the random percolation,
in particular with a correlation length exponent $\nu=2/a$.
In the case of enzyme catalyzed gel proteolysis, the density
of non-hydrolyzed bonds is certainly correlated, and
if the enzyme concentration is small it may be correlated
also at large distances.

\begin{figure}
\bc
a)\mbox{\includegraphics[width=5cm]{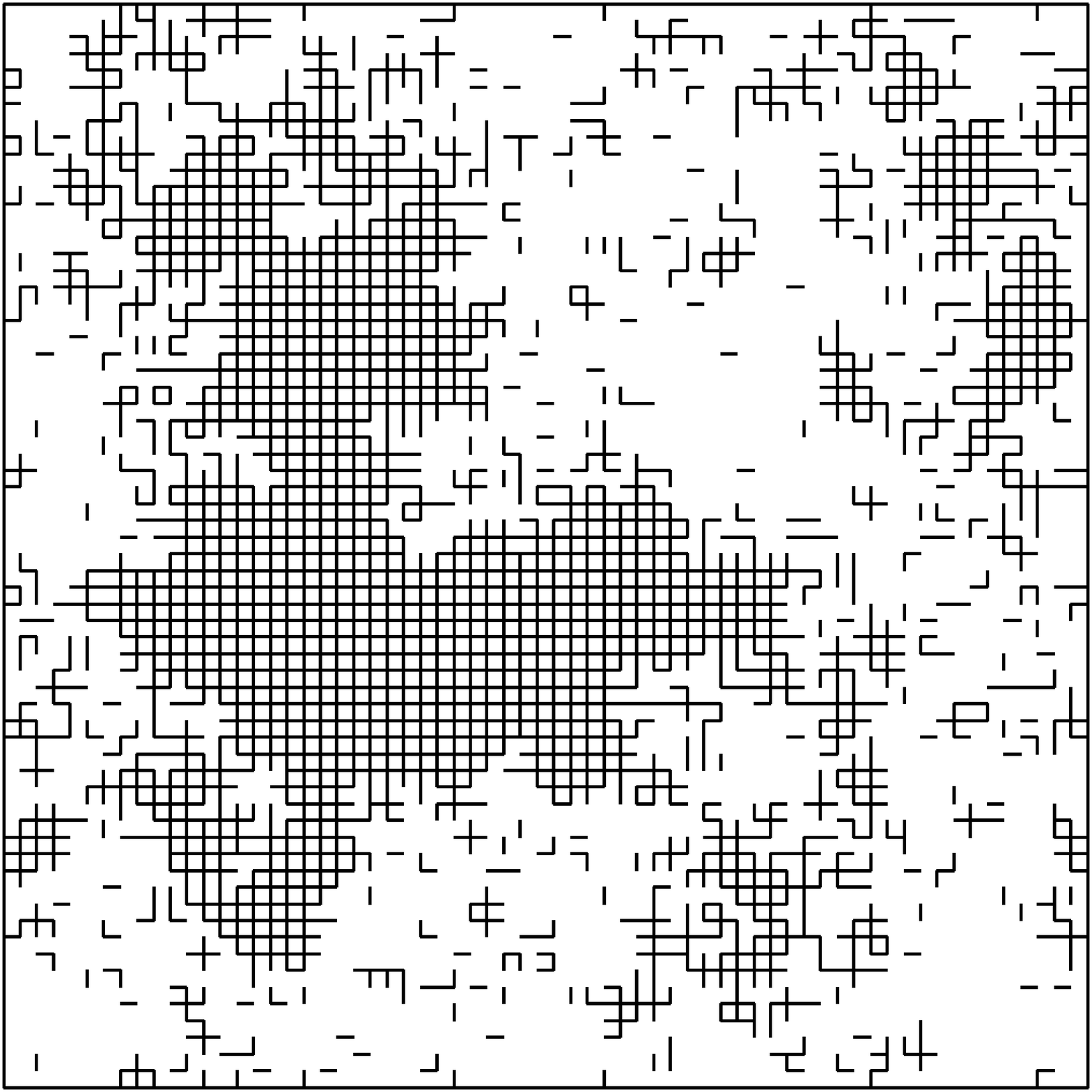}}\\
\bigskip
b)\mbox{\includegraphics[width=5cm]{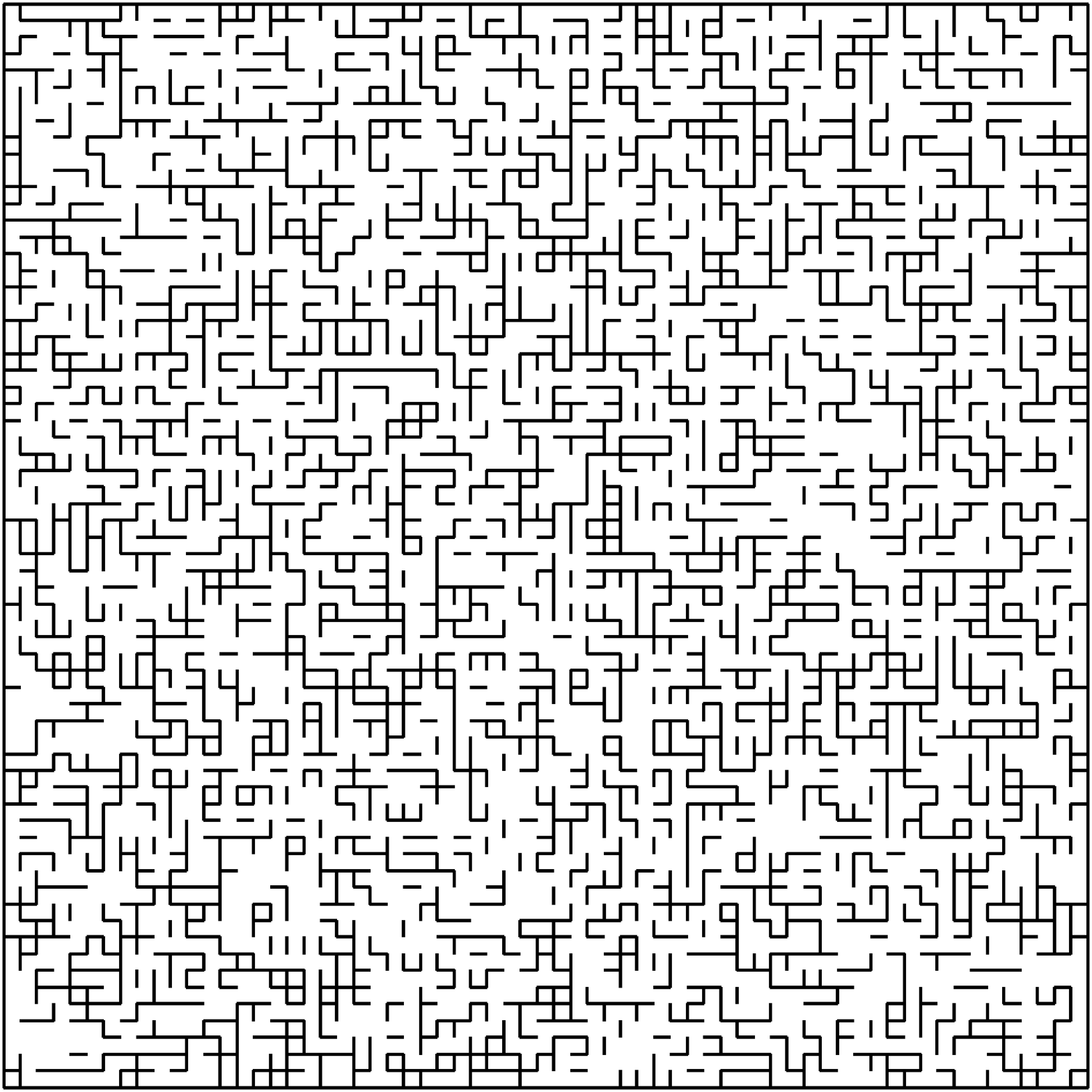}}
\ec
\caption{a) The two-dimensional model of size $64^2$,
after the bond-hydrolyzing enzyme has done $12000$ steps, and
the density of non-hydrolyzed bonds is $p=0.42$.
Bonds are spatially correlated also at large distances.
b) The two-dimensional random percolation model with the same bond density.}
\label{pacman}
\end{figure}

In this paper, we study a very simple model, which we call "{pacman
percolation model}", in which the protein gel is schematized as a cubic
lattice of $N=L^3$ sites, where each site represent an (exavalent)
monomer. At time $t=0$ all the bonds between nearest neighbor monomers
are present. One enzyme is then introduced in the system, that at each
step goes from one site to a nearest neighbor site, chosen randomly
between the six possible neighbors, and hydrolyzes (deletes) the
corresponding bond if not yet hydrolyzed \cite{nota}.
Periodic boundary conditions are chosen.
In Fig.\ \ref{pacman}a it is shown the two-dimensional
version of the model, after the enzyme has walked around for some time
(roughly at the percolation threshold). Note how the remaining
non-hydrolyzed bonds are spatially correlated, with respect to a
random percolation model (Fig.\ \ref{pacman}b).
At each time step, there will be a distribution
of clusters, where two sites belong to the same cluster if
there is a path of non-hydrolyzed bond between them.
We measure then, as a function of the density $p$ of bonds:
a boolean variable equal to one if there is a percolating cluster,
to zero otherwise;
the size of the percolating cluster, if any;
the mean cluster size, that is ${1\over N}\sum_{s} n_s s^2$, where
$n_s$ is the number of clusters of size $s$, and the percolating cluster
is excluded from the sum.
We perform the experiment many times with different random numbers,
and average the above mentioned quantities, obtaining respectively the
percolating probability $\Pi(p,L)$, the percolating cluster density
$\rho(p,L)$, the mean cluster size $\chi(p,L)$, where we have explicited
the dependence on the size $L$ of the lattice.

\begin{table}
\bc
\begin{tabular}{|c|c|c|c|}
\hline
&pacman percolation &experiment &random percolation\\
\hline
$p_c$     & $0.139\pm0.001$ &          & $0.2488$ \\
\hline
$\nu$     & $1.8\pm0.1$     &          & $0.88$   \\
$\beta$   & $1.0\pm0.1$     & $1.0\pm0.1$ & $0.41$   \\
$\gamma$  & $3.4\pm0.2$     &          & $1.80$   \\
$\mu$       & $3.5\pm0.1$     &          & $2.0$    \\
$s$       & $1.1\pm0.1$      &          & $0.73$   \\
\hline
\end{tabular}
\ec
\caption{Percolation density and critical indices
in the pacman percolation model, and in random percolation, in three
dimensions.}
\label{tabella}
\end{table}

\begin{figure}
\bc
\mbox{\includegraphics[width=6cm]{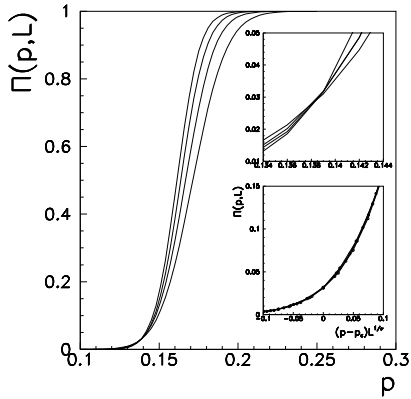}}
\ec
\caption{Percolation probability $\Pi(p,L)$ as a function of the
bond density $p$, for cubic lattices of size $L=30$, 40, 50, 60.
Upper inset: the point of intersection of the curves. Lower inset:
data collapse obtained plotting $\Pi(p,L)$ versus $(p-p_c)L^{1/\nu}$,
with $\nu=1.8$.}
\label{cross}
\end{figure}

\begin{figure}
\mbox{\includegraphics[width=6cm]{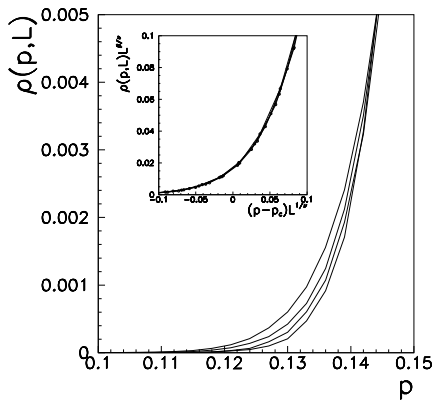}}
\caption{Density $\rho(p,L)$ of the percolating cluster as
a function of the bond density $p$, for the same lattice sizes of
Fig.\ \ref{cross}. Inset: data collapse obtained plotting
$\rho(p,L)L^{\beta/\nu}$ versus $(p-p_c)L^{1/\nu}$,
with $\nu=1.8$ and $\beta=1.0$.}
\label{beta}
\end{figure}

\begin{figure}
\mbox{\includegraphics[width=6cm]{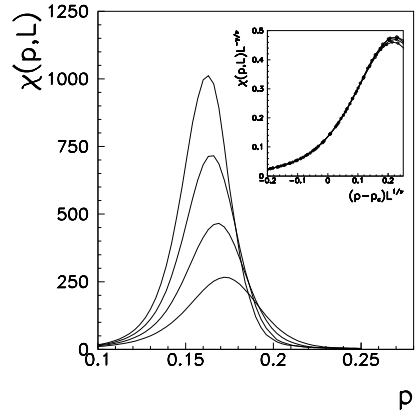}}
\caption{Mean cluster size $\chi(p,L)$ as
a function of the bond density $p$, for the same lattice sizes of
Fig.\ \ref{cross}. Inset: data collapse obtained plotting
$\chi(p,L)L^{-\gamma/\nu}$ versus $(p-p_c)L^{1/\nu}$,
with $\nu=1.8$ and $\beta=3.4$.}
\label{gamma}
\end{figure}

{}From these quantities, it is possible to evaluate the percolation
density $p_c$ and the critical exponents $\nu$, $\beta$ and $\gamma$
\cite{stauffer}. Plotting the percolating probability $\Pi(p,L)$ as a
function of $p$ for different lattice sizes $L$, it is possible to
measure the percolation threshold density $p_c$ as the point in which
the different curves intersect, for $L\to\infty$. In Fig.\ \ref{cross}
it is shown the measured $\Pi(p,L)$ for cubic lattices of size $L=30$,
40, 50, and 60. The intersection point (upper inset) is
$p_c=0.139\pm0.001$. Plotting then $\Pi(p,L)$ as a function of
$(p-p_c)L^{1/\nu}$, one can measure the correlation length exponent
$\nu$ as the value that gives the best collapse of the curves. In the
lower inset of Fig.\ \ref{cross} it is shown the obtained data
collapse, that gives $\nu=1.8\pm0.1$. The confidence interval is
defined by looking when the curves, in the interval shown, do not
collapse anymore within the error bars. In Fig.\ \ref{beta} it is
shown the percolating cluster density for the same system sizes.
Plotting $\rho(p,L)L^{\beta/\nu}$ as a function of $(p-p_c)L^{1/\nu}$, one
can measure the exponent $\beta$ (see inset), We find $\beta=1.0\pm
0.1$, in excellent agreement with the experimental result of Ref.\
\cite{BBA2000}. Finally, in Fig.\ \ref{gamma} we measure the mean
cluster size exponent $\gamma$, finding $\gamma=3.5\pm0.2$. Note that
$\nu$, $\beta$ and $\gamma$ satisfy well the hyperscaling relation
$2\beta+\gamma=\nu d$, expected on general grounds \cite{stauffer}. In
Tab.\ \ref{tabella} the critical exponents found are compared with
those of random percolation. These results show that the model of
``pacman percolation'' falls into another universality class with
respect to random percolation.

\begin{figure}
\mbox{\includegraphics[width=6cm]{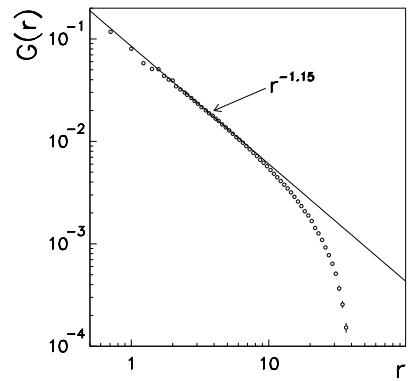}}
\caption{Spatial correlation $G(r)$ in the occupation of the bonds,
for a cubic lattice of size $L=100$.}
\label{correlabo}
\end{figure}

We have also tried to verify the relation predicted by Weinrib and
Halperin \cite{weinrib} between the exponent $\nu$ and the power law
governing the decay of correlations.
In Fig.\ \ref{correlabo} the correlation $G(|i-j|)=\lan n_in_j\ran
-\lan n_i\ran\lan n_j\ran$ in the occupation of the bonds $i$ and $j$, where
$|i-j|$ is the distance between the centers of the bonds, is shown for
a system of size $100^3$ at the percolation threshold $p=0.139$. The
correlation obeys a power law $G(r)\sim r^{-a}$ with $a=1.15 {\pm} 0.05$,
with an exponential cut-off, presumably due to finite size effects, at
distances larger than $r\simeq 30$. The relation $\nu=2/a$ predicted
by Weinrib and Halperin, is quite well verified within the errors. It
has been recently argued \cite {BPJ2003} that for such a model the
correlation between bonds should decay as $1/r$, implying $a=1$ and
$\nu =2$. The prediction, however, is valid only if some conditions
are verified, such as long times and large distances. The discrepancy
between this prediction and our results may be due to the
fact that these asymptotic regimes are not reached in our simulations.

\begin{figure}
\mbox{\includegraphics[width=6cm]{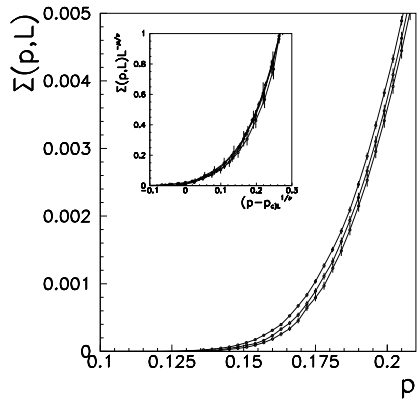}}
\caption{Conductivity of the random-resistor network.}
\label{elastic}
\end{figure}

\begin{figure}
\mbox{\includegraphics[width=6cm]{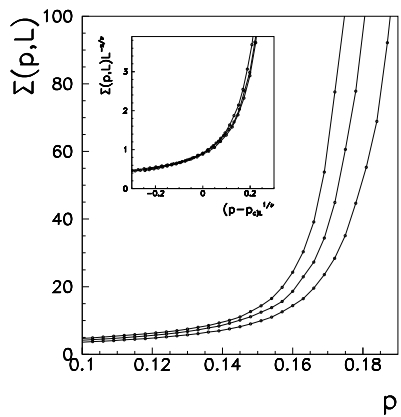}}
\caption{Conductivity of the conductor-superconductor network.}
\label{visco}
\end{figure}

To complete our study, we have analyzed the critical indices
of the conductivity in the random resistor and conductor-superconductor
networks. In the first case each present bond of the model is substituted
with a resistor of unitary conductance, while absent bonds have zero
conductance. The total conductivity $\Sigma$ of the model is then measured
as a function of bond density, and it is zero for $p<p_c$, while it
grows as $|p-p_c|^\mu$ for $p>p_c$. Using finite size scaling as usual
(see Fig.\ \ref{elastic}) we find $\mu=3.5\pm 0.1$.
In the second case each present bond of the model is substituted
with a superconductor of infinite conductance, while absent bonds
are substituted with resistors of unitary conductance. In this case
the total conductivity $\Sigma$ diverges as $|p-p_c|^s$ for $p<p_c$,
and stays infinite for $p>p_c$. In this case we find $s=1.1\pm 0.1$
(see Fig.\ \ref{visco}.)

It was proposed some time ago \cite{deGennes,deGsuperc}, that
the exponents $\mu$ of the resistor network, and $s$ of the
conductor-superconductor network, could be in correspondence respectively
with the exponents $t$ of the elastic modulus, and $k$ of the viscosity.
The first correspondence is based on the analogy between the equilibrium
condition of sites under elastic tension
($\sum_j k_{ij}({\bf r}_i-{\bf r}_j)=0$) and the Kirchhoff equations
for the current conservation ($\sum_j\sigma_{ij}(V_i- V_j)=0$).
However, the electrical problem is a pure scalar
problem, while the elastic one has a vectorial nature, so the
macroscopic elastic response is in principle determined by more
than one elastic modulus.

It is worth to notice that, if the density of enzyme increases, the 
process of degradation of the gel becomes closer to a random 
degradation.
Indeed, in the limit in which each bond is hydrolyzed by a different enzyme,
there will be no correlation between them. This is confirmed by our
simulations. We have analyzed the behavior of the model with a finite 
density $\rho_E$ of enzymes (defined as the ratio between the 
number of enzymes present and the number of sites of the lattice), and 
we have found that for densities greater
than $\rho_E\simeq 0.8$ the system falls in the universality class 
of random percolation. For $\rho_E=0.8$ we find $\nu=0.88\pm 0.02$ and
$\beta=0.41\pm 0.01$, in excellent agreement with random percolation
\cite{stauffer}.

In conclusion, we have used a percolation model to study the
degradation process of ECM due to the action of enzymes. 
Our results shows that, for low density of enzymes, our model
belongs to a different universality class from random percolation. The
change in the critical indices may be due to long range correlation. If the  
density of enzymes is sufficiently high, the correlation between
bonds disappears and there is a crossover to random percolation.

Acknowledgments --
This work was partially supported by INFM-PRA (HOP),
MURST-PRIN-2003, and MIUR-FIRB-2002.

\end{document}